\documentclass[a4paper,11pt]{article}
\usepackage{jheppub} 
\usepackage{amsmath}
\usepackage{physics}
\usepackage{slashed}
\usepackage{graphicx}
\usepackage{xfrac}
\usepackage{nicefrac}
\usepackage{hyperref}
\usepackage{orcidlink}
\usepackage{esvect}
\newcommand\eqdef{\stackrel{\text{def}}{=}}

\arxivnumber{} 

\title{\boldmath Parity anomalies in 2+1-dimensional Aharonov-Bohm type configurations}

\author{H. Basmacı\textsuperscript{\orcidlink{0009-0003-6536-9036}},}
\author{H.O. Cildiroglu \textsuperscript{\orcidlink{0000-0003-0246-1102}}}
\affiliation{Department of Physics Engineering, Ankara University\\
06100, Ankara, Türkiye}

\emailAdd{hoc@physics.bu.edu}

\abstract{We demonstrate that Dirac fermions in 2+1 dimensions, coupled to Abelian gauge fields in multiply-connected regions, exhibit a parity anomaly that directly manifests as Aharonov-Bohm (AB) type topological phases. Using the Fujikawa method, we show that this anomaly reproduces both the AB phase and the spin-dependent Aharonov-Casher (AC) phase. We explicitly calculate the induced topological currents, establishing that their characteristic divergence-free form provides a direct signature of this parity-anomalous topological phase.}

\begin{document}
\maketitle
\flushbottom

\vspace{-0.5cm}

\section{Introduction}
\label{sec:intro}
Topological phases and quantum anomalies exhibit an intimate connection that stems from a common origin in the fundamental aspects of quantum theory \cite{Berry1984, Jackiw1985, Nelson1985, Stone1986, Fujikawa2006, Witten2015, Gamboa2025}. Establishing a unified framework for these phenomena, particularly between Aharonov-Bohm (AB) type effects and parity anomalies, provides deep physical insights \cite{Bracken2008}. This connection manifests profoundly in 2+1 dimensions, where their underlying characteristics become strikingly apparent.

The fundamental nature of AB effects inherently resides in three spatial dimensions. However, the system's invariance under spatial translations along the $z$-direction allows the system to be confined to an effective 2+1-dimensional description. This framework reduces the problem to analyzing particles moving along closed trajectories around a topological singularity, treated as an effective electric/magnetic monopole or polarized dipole on the plane \cite{Cildiroglu2024}. These interactions, originating from gauge potentials and manifesting without classical forces, result in the well-known AB phase for electrons moving around magnetic dipoles and the Aharonov-Casher (AC) phase for magnetic dipoles moving around electric charges \cite{Aharonov1959, Chambers1960, Tonomura1986, Peshkin1989, Aharonov1984, Cimmino1989}.
 
In 2+1 dimensions, the classical parity symmetry of massless Dirac fermions is quantum-mechanically broken upon coupling to a background gauge field, thereby inducing the parity anomaly. This anomaly persists for massive fermions, where the Dirac mass explicitly breaks parity at the classical level, yet quantum effects induce analogous topological responses, resulting in a current perpendicular to electromagnetic fields \cite{Stone2023, Niemi1983, Redlich1984, Deser1982, Korchemsky1991}. The resulting anomaly-induced topological current is mass-independent at leading order, with any mass dependence confined to subleading corrections. The mathematical foundation of this phenomenon lies in the Atiyah-Patodi-Singer (APS) index theorem, which describes the spectral properties of Dirac operators in odd-dimensional manifolds with boundaries \cite{APS1975, AS1963}. Therefore, conventional AB-type setups constitute ideal platforms for analyzing the parity anomaly through the direct measurement of the topological current.

Here, we demonstrate that Dirac fermions interacting with Abelian gauge fields in multiply-connected regions induce the parity anomaly in AB-type setups. Using the Fujikawa path integral method \cite{Fujikawa1979, Fujikawa1980, Fujikawa1980err, Fujikawa2004}, we show this anomaly manifests directly as a complex phase factor in the particle wave functions. We further established that the parity anomaly generates a planar induced topological current, which is tied to the AB phases.

This paper is organized as follows: In the second section, we provide a review of parity anomalies in 2+1 dimensions, applying the Fujikawa method to AB and AC configurations. In the third section, we calculate the induced topological currents from the variation of the partition function over the electromagnetic potentials. Lastly, we discuss the results and possible directions for future work.

\section{Parity Anomalies}
\label{sec:par}

\subsection{Aharonov-Bohm Case}

To analyze the parity anomaly in the AB configuration, we begin with the relativistic Lagrangian for an electron in electromagnetic fields:
\begin{align}
    \mathcal{L}_{\text{AB}} = \bar{\psi}\left(i\gamma^{\mu}\partial_{\mu}+e\gamma^{\mu} A_{\mu} - m \right)\psi.
    \label{eq:2.1}
\end{align}
In 2+1 dimensions $(+, -, -)$, the gamma matrices satisfy the Clifford algebra in two distinct representations,
\begin{equation}
\begin{matrix}        
\alpha_x=\sigma_x \\
\alpha_y=s \sigma_y \\
\beta=\gamma_0=\sigma_z
\end{matrix}
\hspace{5mm}
\begin{matrix}        
\gamma_1=i\sigma_y \\
\gamma_2=-is\sigma_x
\end{matrix}
\label{eq:2.2}
\end{equation}
where $s=\pm 1$ denotes the spin polarizations. The antisymmetric tensor $\sigma^{\mu\nu}$ is given by
\begin{align}
    \sigma^{\mu \nu} = \frac{i}{2}\left[\gamma^{\mu}, \gamma^{\nu}\right],
\label{eq:2.3}
\end{align}
\noindent leading to the identities $\gamma^0 \gamma^i = \alpha^i$, $\sigma^{0i} = i \alpha^i$, and $\sigma^{ij} = \sigma^{k}$. Hence, the term $\gamma^{\mu}A_{\mu}$ in Eq. \eqref{eq:2.1} needs to be redefined. In the AB configuration, the scalar potential is zero ($\phi = A_0 = 0$), and the planar confinement ideally eliminates any $z$-component of the vector potential ($A_3 = 0$). The four-potential $A_{\mu} = (0, A_1, A_2, 0)$ results in $\gamma^{\mu}A_{\mu} = \gamma^1A_1 + \gamma^2A_2 = \gamma^iA_i$. By separating the time and spatial components, the Lagrangian in \eqref{eq:2.1} is written as:
\begin{align}
    \mathcal{L}_{\text{AB}} & = \bar{\psi}\left(i\gamma^0\partial_0 + i\gamma^i\partial_i + e\gamma^iA_i - m\right)\psi. 
    \label{eq:2.4}
\end{align}
This expression takes the standard Lagrangian form $\mathcal{L}_{D}=\bar{\psi}\left(\slashed{D}-m \right)\psi$, where the covariant derivative of the AB Lagrangian $\slashed{D} \eqdef \gamma^{\mu}D_{\mu}$ comprises $\slashed{D} = i\gamma^0\partial_0 + i\gamma^i\partial_i + e\gamma^iA_i$. We identify the temporal and spatial parts of the covariant derivative $\slashed{D}_t = i\gamma^0\partial_0$ and $\slashed{D}_s = i\gamma^i\partial_i + e\gamma^iA_i$, respectively. 

The quantum dynamics are fully encoded in the partition function $(Z)$ constructed from $\mathcal{L}_{\text{AB}}$,
\begin{align}
    Z = \int \prod [\dd \psi \, \dd \bar{\psi}] \, e^{-\int \mathcal{L}_{\text{AB}} \, \dd x \, \dd t}. 
\label{eq:2.5}
\end{align}
The spinor fields $\psi$ and $\bar{\psi}$ can be decomposed by expanding in the eigenfunctions of $\slashed{D}_s$ and $\slashed{D}_s^{\dagger}$,
\begin{align}
    \psi(\textbf{x}, t) & = \sum_{n} a_n(t)\phi_n(\textbf{x}) \nonumber \\
    \bar{\psi}(\textbf{x}, t) & = \sum_{m} b_m(t)\phi_m^{\dagger}(\textbf{x})
\label{eq:2.6}
\end{align}
Thus, the instantaneous eigenvalue equations are defined as,
\begin{align}
\slashed{D}_s\phi_n(\textbf{x}) & = \left[\gamma^i(i\partial_i + eA_i) \right]\phi_n(\textbf{x}) = i\lambda_n\phi_n(\textbf{x}) \nonumber \\ 
\phi_n^{\dagger}(\textbf{x})\slashed{D}_s^{\dagger} & = \phi_n^{\dagger}(\textbf{x})\left[\gamma^i(i\, \overleftarrow{\partial_i} + eA_i) \right] = i\lambda_n\phi_n^{\dagger}(\textbf{x})
\label{eq:2.7}
\end{align}
Here, $\slashed{D}_s^\dagger = \slashed{D}_s$ establishes the Hermiticity of the spatial covariant derivative, implying real eigenvalues $\lambda_n$ considering the direction of the derivative. Substituting \eqref{eq:2.6} and \eqref{eq:2.7} into \eqref{eq:2.4}, the AB Lagrangian becomes,
\begin{align}
    \mathcal{L}_{\text{AB}} =  i&b_m\phi_m^{\dagger}\gamma^0\partial_0a_n\phi_n + i\lambda_na_nb_m\phi_m^{\dagger}\phi_n - ma_nb_m\phi_m^{\dagger}\phi_n.
    \label{eq:2.8}
\end{align}
To perform the parity transformation, we use $\gamma^0$ as a generator \cite{Dittrich1986}. While the classical Lagrangian remains invariant under the  $\psi \to e^{i\gamma^0 \alpha(x)}\psi$ transformation, $Z$ in \eqref{eq:2.5} acquires a non-trivial Jacobian, where $\alpha(x)$ is a gauge function. Regularization via Fujikawa's method \cite{Fujikawa1979, Fujikawa1980, Fujikawa1980err, Fujikawa2004} yields the anomaly term $\mathcal{M}$. Accordingly, spatial integration of Eq. \eqref{eq:2.8} leads to the partition function $Z$ expressed as a path integral over the expansion coefficients. The parity transformation then induces factorization $Z = Z_0Z'$, separating the time-independent $(Z_0)$ and time-dependent $(Z')$ sectors. The determinant of the Jacobian becomes the anomaly term after regularization. Using a momentum cutoff $M$, the anomaly term $\mathcal{M}$ is expressed as a trace over the complete set of basis functions,
\begin{align}
    \mathcal{M} = \lim_{M \to \infty} \int \dd^2 x \, \alpha\left(x\right) \int \frac{\dd^2 k}{\left(2\pi\right)^2} \, \Tr\left[e^{-i k \cdot x} \gamma^0 e^{\nicefrac{-\slashed{D}^2}{M^2}} e^{i k \cdot x}\right].
    \label{eq:2.9}
\end{align}

For the AB case, with the identities $\{ \gamma_{\mu}, \gamma_{\nu} \} = 2\eta_{\mu \nu}$ and $[D^{\mu}, D^{\nu}] = ieF^{\mu \nu}$, where $D^{\mu} = \partial^{\mu} + ieA^{\mu}$, the operator $\slashed{D}^2 = \gamma_\mu\gamma_u D^\mu D^u$ can be rearranged as,
\begin{align}
    \slashed{D}^2 & = D^{\mu}D_{\mu} + \frac{1}{2}ie\gamma_{\mu}\gamma_{\nu}F^{\mu \nu}.
    \label{eq:2.10}
\end{align}
Here, $F^{\mu \nu}$ is the electromagnetic field tensor, which can be redefined in 2+1 dimensions, using $F_{\mu \nu} = \eta_{\mu \alpha} F^{\alpha \beta} \eta_{\beta \nu}$ as in the matrix form,
\begin{align}
    F_{\mu \nu} =
    \begin{pmatrix}
        0 & E_x & E_y \\
        -E_x & 0 & -B \\
        -E_y & B & 0 \\
    \end{pmatrix}.
    \label{eq:2.11}
\end{align}

Then, expanding the exponential term $e^{\nicefrac{-\slashed{D}^2}{M^2}}$ in powers of $M^{-2}$ up to $\mathcal{O}(M^{-2})$ and substituting $\slashed{D}^2$ into Eq. \eqref{eq:2.9} yields the following form, as higher-order terms vanish in the limit $M \to \infty$,
\begin{align}
\mathcal{M}=\lim_{M \to \infty}& \int \dd^2 x \ \alpha\left(x\right) \int \frac{\dd^2 k}{\left(2\pi\right)^2} \ e^{\nicefrac{-(k + eA)^2}{M^2}} \Tr\left[\gamma^0 \left(1 - \frac{1}{2M^2} ie \gamma_{\mu} \gamma_{\nu} F^{\mu \nu}\right) \right]
\label{eq:2.12}
\end{align}
Lastly, the anomaly term is obtained after evaluating the momentum integral and utilizing the trace identities $\Tr[\gamma^0] = 0$ and $\Tr[\gamma^{\mu} \gamma^{\nu} \gamma^{\rho}] = -2i\epsilon^{\mu \nu \rho}$ as:
\begin{align}
    \mathcal{M} = -\frac{e}{4\pi}\int \dd^2 x \, \alpha\left(x \right) \epsilon_{0 \mu \nu} F^{\mu \nu}.
    \label{eq:2.13}
\end{align}
\noindent The anomaly term in Eq. \eqref{eq:2.13} takes its sole contribution from the spatial indices $\mu=1$, $\nu=2$ (or equivalently $\mu=2$, $\nu=1$), as all other components vanish. Substituting $F^{12} = -B$ from \eqref{eq:2.11} and selecting $\alpha(x) = 4\pi$ yields the anomaly term by recalling that the closed line integral of the vector
potential is the magnetic flux $\oint A \cdot \dd l = \Phi$:
\begin{align}
\mathcal{M} = e \Phi.
\label{eq:2.14}
\end{align}
Accordingly, the anomaly term $\mathcal{M}$ emerges in the form of the familiar AB phase in the current configuration. In other words, the parity anomaly enters the wave function of the particles as the topological phase, leading to observable consequences. In \eqref{eq:2.14}, the anomaly term is independent of the spin orientation of the particles, as it does not explicitly involve the variable $s$. Furthermore, it arises independently of the particle velocities and the geometry of the path, depending solely on the topology of the system, which is consistent with the nature of the AB effect.

\subsection{Aharonov-Casher Case}
In the AC effect, the topological phase arises from the interaction of magnetic dipole moment $(\mu)$ of a neutral particle with the electric field generated by an infinitely thin and long linear charge distribution $(\lambda)$ \cite{Aharonov1984, Cimmino1989}. This interaction introduces a complex phase contribution to the particle's wave function in the absence of classical forces. The established interaction highlights a profound electric/magnetic dualities between the AC and AB phases, which becomes strikingly apparent in 2+1 dimensions. The linear charge distribution is conceptually viewed as a coherent array of electrons, effectively becoming a charged particle when confined on a plane. The problem thereby simplifies to the motion of an unpolarized magnetic dipole in a closed orbit around an electron. The AC effect represents the more general phenomenon that arises in the stationary reference frame of the AB electrons, describing the case of unpolarized dipoles, which allows their spin orientation to be tracked. Thus, the AC configuration provides a broader framework for probing the parity anomaly through topological phases, where the resulting anomaly term is also expected to depend on the spin polarizations of moving dipoles as an additional degree of freedom.

In this regard, we consider the relativistic Lagrangian describing the dynamics of a point-like particle carrying an electric charge with a spin-carrying neutral object, consistent with the existing configuration:
\begin{align}
    \mathcal{L}^{\pm}_{\text{AC}} = \bar{\psi}^{\pm}\left(i\gamma^{\mu}\partial_{\mu}+\frac{\mu}{2}\sigma^{\mu \nu}F_{\mu \nu} - m\right)\psi^{\pm}.
    \label{eq:2.15}
\end{align}
\noindent Here, polarized spinors emerge through spin projection operators acting on their unpolarized counterparts \cite{Singleton2016}:
\begin{align}
    \psi^{\pm} & = \Sigma^{\pm} \psi \, \, \, \, \text{and} \, \, \, \, \bar{\psi}^{\pm} = \Sigma^{\pm}\bar{\psi} \nonumber \\ & \Sigma^{\pm} = \frac{1}{2} \left(1 \pm \frac{1+s}{2} \sigma_z\right)
    \label{eq:2.16}
\end{align}

In 2+1 dimensions, the planar confinement necessitates a redefinition of the interaction term $\sigma^{\mu \nu}F_{\mu \nu}$. For the AC configuration, the magnetic field vanishes $(B = 0)$. Combining this with the antisymmetric tensor properties, we obtain
\begin{align}
    \sigma^{\mu \nu}F_{\mu \nu} = 2\sigma^{0i}F_{0i} + \sigma^{ij}F_{ij} = 2i\alpha^iE_i.
    \label{eq:2.17}
\end{align}
Since the term given by the Lagrangian \eqref{eq:2.15} has components defined in a two-dimensional space, the three-dimensional electric field vector must be confined to a plane $(\tilde{E} = E\times \hat{z})$. Therefore, the two-dimensional vector $\tilde{E}$ has components $(E_y, -E_x, 0)$, and \eqref{eq:2.17} becomes $s\mu \gamma^{i}\tilde{E}_i$. Now, leveraging the idempotency $\Sigma^{\pm}\Sigma^{\pm} = \Sigma^{\pm}$ of the projection operators, we decompose the term into its kinetic and dipole coupling components:
\begin{align}
    \Sigma^{\pm} \left(i\gamma^{\mu}\partial_{\mu}\right) \Sigma^{\pm} & = \frac{i}{2} \left(\gamma^{\mu} \partial_{\mu} \pm \frac{1+s}{2} \gamma^0 \gamma^{\mu} \partial_{\mu} \right) \nonumber \\
    \Sigma^{\pm} (s\mu \gamma^i \tilde{E}_i) \Sigma^{\pm} & = \frac{\mu}{2} \left(s\gamma^i \tilde{E}_i \pm \frac{1+s}{2}\gamma^0 \gamma^i \tilde{E}_i\right)
    \label{eq:2.18}
\end{align}
These expressions can be combined to yield the Dirac operator:
\begin{align}
    \slashed{D}=&\frac{i}{2}\left(\gamma^{\mu} \partial_{\mu} \pm \frac{1+s}{2} \gamma^0 \gamma^{\mu} \partial_{\mu} \right) + \frac{\mu}{2} \left(s\gamma^i \tilde{E}_i \pm \frac{1+s}{2}\gamma^0 \gamma^i \tilde{E}_i\right).
    \label{eq:2.19}
\end{align}
Following the eigenfunction expansion of the spinor fields, analogous to Eqs. \eqref{eq:2.6} and \eqref{eq:2.7}, the Lagrangian takes the form of \eqref{eq:2.8}. This structural analogy implies that the parity transformation procedure is directly applicable. Hence, the partition function acquires a non-trivial Jacobian under the transformation $\psi \to e^{i \gamma^0 \alpha(x)} \psi$. We evaluate the Jacobian via Fujikawa regularization with a cutoff $M$. After expanding the exponential to $\mathcal{O}(M^{-2})$ and inserting $\slashed{D}^2$, we proceed with spatial integration and factorization of the partition function, which reveals the anomaly:
\begin{align}
    \mathcal{M} = - \lim_{M \to \infty} \frac{1}{M^2} &\int \dd^2 x \, \alpha\left(x\right) \int \frac{\dd^2 k}{\left(2\pi\right)^2} \, \Tr\left[e^{-i k \cdot x} \gamma^0 \slashed{D}^2 e^{i k \cdot x}\right].
    \label{eq:2.20}
\end{align}
Using the identities $\Tr[\gamma^0 \gamma^{\mu} \gamma^{0} \gamma^{\nu}] = 4 \eta^{0 \mu} \eta^{0 \nu} - 2\eta^{\mu \nu}$ and $\Tr[\gamma^{\mu}\gamma^{\nu}] = 2\eta^{\mu \nu}$, the trace in \eqref{eq:2.20} is rearranged as $\Tr\left[e^{-i k \cdot x} \gamma^0 \slashed{D}^2 e^{i k \cdot x}\right] = \frac{s\mu}{2} \, [\nabla \cdot E]$. Last, the momentum integration is resolved by the regularization:
\begin{align}
    \mathcal{M} = - \frac{s\mu}{8\pi} \int \dd^2 x \, \alpha\left(x\right) \, [\nabla \cdot E].
    \label{eq:2.21}
\end{align}
With conventional normalization $\alpha(x) = -8\pi$, the surface integral of the electric field divergence can be identified as the linear charge density $\delta_{\text{AC}} = s\mu \int \left(\nabla \cdot E\right) \, \dd a = s\mu \lambda$. The anomaly then reduces to
\begin{align}
    \mathcal{M} = s\mu \lambda.
    \label{eq:2.22}
\end{align}
Here, the derived anomaly term is thus identified as the topological AC phase, which explicitly depends on the spin polarization. This fundamentally distinguishes the AC anomaly from its spin-independent AB counterpart. Nevertheless, the anomaly retains the essential topological property of independence from particle kinematics and path geometry, depending only on the system's topology.

\section{Induced Currents}
\label{sec:ind}

\subsection{Aharonov-Bohm Case}
Coupling a quantum system to an external electromagnetic field generates induced currents, which characterize the system's electromagnetic response \cite{Fujikawa2004, Fradkin2013}. These currents encode fundamental topological and magnetic properties of the underlying quantum state and arise from the variation of the partition function with respect to the applied field. The emergence of these currents can be understood by analyzing how the partition function responds to small (adiabatic) changes in the electromagnetic field. The variation of the partition function can be analyzed by expressing it as a determinant involving the Dirac operator $Z = \det(\slashed{D}+m)$,
where $\slashed{D}$ is the Euclidean Dirac operator that satisfies the eigenvalue equation,
\begin{align}
\slashed{D}\phi_n = i \lambda_n \phi_n,
\label{eq:3.1}
\end{align}
where $(\slashed{D} + m)_{lm} = (i\lambda_l + m)\delta_{lm}$ in terms of the complete set of eigenfunctions $\phi_n$. Under the transformations of electromagnetic potentials $A_{\mu} \to A_{\mu} + \delta A_{\mu}$, one can calculate the variation of the logarithm of the partition function using standard functional methods \cite{Fujikawa2004, Fradkin2013}. The induced current density is then given by
\begin{align}
    j_{\rho} = \frac{\delta \ln{Z}}{\delta A_{\rho}}.
    \label{eq:3.2}
\end{align}
Using the completeness relations in momentum space, the variation yields,
\begin{align}
\delta \ln{Z} = -e  &\int_0^{\infty} \dd s \, e^{-m^2s} \int_{-\infty}^{\infty} \frac{\dd^3 k}{(2\pi)^3} e^{i k \cdot x} \Tr[(\slashed{D}-m)e^{\slashed{D}^2 s} \gamma^{\mu} \delta A_{\mu}] e^{i k \cdot x}.
\label{eq:3.3}
\end{align}
We now use the definition \eqref{eq:2.10} and expand the exponential in the coupling constant. Since integrals of the form $\int \dd k_{\mu} \, k_\mu e^{-k^2 s} = 0$ vanish by symmetry, the expression becomes
\begin{align}
\delta \ln{Z} = \frac{m e^2}{8 \pi}& \int_0^{\infty} \dd s \, e^{-m^2s} \int_{-\infty}^{\infty} \frac{\dd k_0}{2\pi} e^{-(k_0 + e A_0)^2s} \Tr[\gamma_{\mu}\gamma_{\nu}\gamma_{\rho}] F^{\mu \nu} \delta A^{\rho}.
\label{eq:3.4}
\end{align}
The spatial momentum integrations can be performed analytically, since all the $k$-dependence is contained in the exponential factor. Applying the trace identities and performing the remaining integrations over proper time $s$ and temporal momentum $k_0$, yields the variation: $\delta \ln{Z} = \frac{e^2}{8\pi} \epsilon_{\mu \nu \rho} F^{\mu \nu} \delta A^{\rho}$. From the definition of the current \eqref{eq:3.2}, it gives the induced current density:
\begin{align}
j_{\rho} = \frac{e^2}{8\pi} \epsilon_{\mu \nu \rho} F^{\mu \nu}.
\label{eq:3.5}
\end{align}

Thus, the induced current is determined by the electromagnetic field tensor and exhibits the characteristic form of the parity anomaly in Eq. \eqref{eq:2.13}. It indicates that the AB interaction breaks the parity symmetry and leads to a parity anomaly. Furthermore, employing the Bianchi identity reveals that $j_{\rho}$ is divergence-less:
\begin{align}
    \partial^{\rho}j_{\rho} = 0,
    \label{eq:3.6}
\end{align}
which reflects the non-locality of the AB effect, that is, $\delta_{AB}$ depends on the total magnetic flux rather than a local field.

\subsection{Aharonov-Casher Case}
The calculation for the AC setup proceeds analogously, with the key modification that the coupling constant transforms as $e \to \nicefrac{\mu}{2}$, reflecting the replacement of electric charge by magnetic dipole moment. This alters the squared Dirac operator to:
\begin{align}
\slashed{D}^2 = D^{\mu} D_{\mu} + \frac{\mu}{4} \sigma_{\mu \nu} F^{\mu \nu}.
\label{eq:3.7}
\end{align}
Substituting \eqref{eq:3.7} into the variation and evaluating the trace with $\Tr[\sigma_{\mu \nu} \gamma_{\rho}] = 2 \epsilon_{\mu \nu \rho}$, we obtain:
\begin{align}
\delta \ln{Z} = \frac{m \mu^2}{32 \pi} \int_0^{\infty} \dd s \, e^{-m^2s} \int_{-\infty}^{\infty} \frac{\dd k_0}{2\pi} \, e^{-(k_0 + \frac{\mu}{2} A_0)^2s} \Tr[\sigma_{\mu \nu} \gamma_{\rho}] \, F^{\mu \nu} \delta A^{\rho}.
\label{eq:3.8}
\end{align}
Performing the integrations over $s$ and $k_0$ yields:
\begin{align}
\delta \ln{Z} = - \frac{\mu^2}{32\pi} \epsilon_{\mu \nu \rho} F^{\mu \nu} \delta A^{\rho}.
\label{eq:3.9}
\end{align}
The induced current density for the AC configuration then follows from \eqref{eq:3.2}:
\begin{align}
j_{\rho} = - \frac{\mu^2}{32\pi} \epsilon_{\mu \nu \rho} F^{\mu \nu}.
\label{eq:3.10}
\end{align}
This expression mirrors the mathematical structure of the AB case, distinguished by the opposite sign and the coupling strength $\mu^2$ in place of $e^2$. The divergence-free character $\partial^{\rho}j_{\rho} = 0$ is again guaranteed by the Bianchi identity, confirming that the AC phase depends on the total linear charge density rather than local field configurations. This establishes the dual topological nature of the AC effect relative to its AB counterpart \cite{Cildiroglu2021}.

\section{Conclusion}
\label{sec:conc}

In this work, we establish the parity anomaly in 2+1 dimensions as the quantum field theoretical origin of AB-type topological phases. By systematically applying the Fujikawa path integral method, we derive anomaly terms that precisely reproduce the AB phase $\mathcal{M}=e\Phi$ and the AC phase $\mathcal{M} = s\mu \lambda$. While both anomalies share a common origin in the multiply-connected topology, our results demonstrate that the AC case is fundamentally distinguished by its explicit dependence on spin polarization. This introduces an additional degree of freedom absent in the AB configuration. Nevertheless, both anomalies remain independent of particle velocities and path geometry, determined exclusively by the topology of the system.

The topological nature of these phenomena is further illuminated by the induced vacuum currents derived from the partition function. We obtain divergence-free current densities for both configurations, confirming that the electromagnetic response is governed by global topological invariants rather than local field dynamics. Here, the currents exhibit an intrinsic duality—manifested in the interchange of coupling constants and sign reversal—reflecting the symmetry between charge-flux and dipole-charge interactions. This framework aligns with the broader context of geometric phases, suggesting that such phases can be systematically understood through the non-trivial Jacobian of the path integral measure in multiply-connected spaces.

Our framework suggests a natural extension to the broader family of AB-type effects, particularly the He-McKellar-Wilkens (HMW) \cite{He1993, Wilkens1994, Gillot2013} and Dual Aharonov-Bohm (DAB) phases \cite{Dowling1999, Cildiroglu2019}. Given the duality relations linking these phenomena \cite{Cildiroglu2021}, they are expected to exhibit commensurate anomaly structures within the 2+1-dimensional formulation. Systematic investigation of these cases, along with their time-dependent generalizations \cite{Singleton2013, Bright2015, Jing2017, Ma2017, Choudhury2019, Jing2020}, constitutes a natural trajectory for future inquiry.

\acknowledgments
The authors would like to express their gratitude to Prof. Müge Boz and the late Prof. Namık Kemal Pak, whose vision and guidance continue to illuminate the path of this research. The authors are also grateful to Prof. Jorge Gamboa for his valuable contributions and insightful discussions, and to Prof. Francisco Correa for the research opportunities he provided during H. Basmacı's visiting researcher period at USACH.




\end{document}